\documentclass[iop,apj]{emulateapj}
\usepackage{apjfonts}
\usepackage{multirow}
\usepackage{natbib}

\def\aj{{AJ}}

\def\apj{{ApJ}}
\def\apjs{{ApJS}}
\def\asec{$^{\prime\prime}$}

\def\deg{$^{\circ}$}

\def\ha{H$\alpha$}
\def\hb{H$\beta$}

\def\kms{km s$^{-1}$}
\def\lamb{$\lambda$}
\def\lax{\mathrel{\hbox{\rlap{\hbox{\lower4pt\hbox{$\sim$}}}\hbox{$<$}}}}
\def\gax{$\mathrel{\hbox{\rlap{\hbox{\lower4pt\hbox{$\sim$}}}\hbox{$>$}}}$}
\def\simlt{\lower.5ex\hbox{$\; \buildrel < \over \sim \;$}}
\def\simgt{\lower.5ex\hbox{$\; \buildrel > \over \sim \;$}}

\def\mbh{{$M_{\rm BH}$}}

\def\mnras{{MNRAS}}
\def\nat{{Nature}}
\def\pasp{{PASP}}

\def\percm2{cm$^{-2}$}

\def\solmass{$M_\odot$}

\def\oiii{[\ion{O}{3}]}

\def\nii{[\ion{N}{2}]}

\def\sii{[\ion{S}{2}]}

\def\farcs{\hbox{$.\mkern-4mu^{\prime\prime}$}}

\shorttitle{ULX in NGC 5252}

\begin{document}

\title{Ionized Gas Kinematics around an Ultra-luminous X-ray Source in 
NGC 5252 : Additional Evidence for an Off-nuclear AGN}

\author{Minjin Kim\altaffilmark{1,2}, 
Luis C. Ho\altaffilmark{3,4},
and Myungshin Im\altaffilmark{5} 
}

\altaffiltext{1}{Korea Astronomy and Space Science Institute, Daejeon 305-348, 
Republic of Korea}

\altaffiltext{2}{University of Science and Technology, Daejeon 305-350, 
Republic of Korea}

\altaffiltext{3}{Kavli Institute for Astronomy and Astrophysics, 
Peking University, Beijing 100871, China} 

\altaffiltext{4}{Department of Astronomy, School of Physics, 
Peking University, Beijing 100871, China} 

\altaffiltext{5}{
Center for the Exploration of the Origin of the Universe (CEOU), 
Astronomy Program, Department of Physics and Astronomy, 
Seoul National University, 599 Gwanak-ro, Gwanak-gu, Seoul, 
151-742, Republic of Korea}

\begin{abstract}
The Seyfert 2 galaxy NGC 5252 contains a recently identified ultra-luminous 
X-ray (ULX) source that has been suggested to be a possible candidate 
off-nuclear low-mass active galactic nucleus.  We present follow-up optical 
integral-field unit observations obtained using GMOS on the Gemini-North 
telescope. In addition to confirming that the ionized gas in the vicinity of 
the ULX is kinematically associated with NGC 5252, the new observations reveal
ordered motions consistent with rotation around the ULX.   The close 
coincidence of the excitation source of the line-emitting gas with the position
of the ULX further suggests that ULX itself is directly responsible for the 
ionization of the gas.  The spatially resolved measurements of 
\nii\ \lamb6584/\ha\ surrounding the ULX indicate a low gas-phase metallicity,
consistent with those of other known low-mass active galaxies but not that of 
its more massive host galaxy.  These findings strengthen the proposition 
that the ULX is not a background source, but rather that it is the nucleus of 
a small, low-mass galaxy accreted by NGC 5252.
\end{abstract}


\section{Introduction}
Most massive galaxies harbor central supermassive black holes (BHs) with \mbh\ 
$\geq 10^{5-6}$ \solmass. BHs with \mbh\ $> 10^9$ \solmass\ have even been 
discovered in high redshift ($z\approx 6-7$) quasars 
(e.g., \citealt{mortlock_2011}; \citealt{jun_2015}; \citealt{wu_2015}).  This 
finding suggests that the seeds of supermassive BHs were probably not ordinary 
stellar-mass BHs (\mbh\ $\leq 20$ \solmass).  Instead, theoretical studies 
suggest that more likely candidates for quasar seeds may be intermediate-mass 
BHs (IMBHs; $10^2$ \solmass $< M_{\rm BH} < 10^5 $\solmass), which grow into 
supermassive BHs through rapid accretion and cosmological merging 
(\citealt{volonteri_2003}).  Despite their importance, IMBHs are surprisingly 
rare in present-day galaxies (e.g., \citealt{greene_2004}).  Thus, the 
systematic study of the demography of IMBHs is crucial to understand the link 
between stellar-mass BHs and supermassive BHs (\citealt{volonteri_2010}).

In this regard, ultraluminous X-ray sources (ULXs) are intriguing because
some ULXs may contain IMBHs.  ULXs are defined as bright, off-nuclear point 
sources with X-ray luminosities $L_{\rm X} > 10^{39}$ erg s$^{-1}$. Because 
their X-ray luminosities substantially exceed the Eddington luminosity of 
stellar-mass BHs, and they are off-nuclear, the X-rays are thought to 
originate from IMBHs (\citealt{colbert_1999}).  However, due to the fact that 
the optical counterparts of ULXs are usually unknown or very faint, little is 
known about the true nature of ULXs (e.g., \citealt{roberts_2008}; 
\citealt{gladstone_2013}).

\citet{kim_2015} newly discovered a candidate IMBH associated with a ULX. This 
ULX is seen at a projected separation of 22\asec\ ($\sim 10$ kpc) from the 
nucleus of the type 2 Seyfert S0 galaxy NGC 5252.  Its X-ray luminosity of 
$\sim 1.2 \times 10^{40}$ erg s$^{-1}$ exceeds the upper limit of stellar-mass 
BHs (e.g., high-mass X-ray binaries; \citealt{swartz_2011}).  
A longslit optical 
spectrum of the ULX obtained with the Magellan/Baade 6.5 meter telescope 
reveals strong emission lines with relatively small systematic velocity offset 
($\sim 13$ \kms) with respect to the nucleus of the galaxy.
In conjunction with multiwavelength spectral energy distribution from X-ray to 
radio energies, Kim et et al. suggest that the ULX in NGC 5252 can be 
explained by an IMBH with $M_{\rm BH} \approx 10^{3-9}$ \solmass.  While the 
BH mass estimate is highly uncertain, this object constitutes one of the 
unique examples of an ``off-nucleus'' active galactic nucleus (AGN) with an 
IMBH.  

This paper presents an optical integral-field spectrum of the ULX to further
explore the nature of the ULX in NGC 5252.  We find that the ionized gas 
surrounding the ULX is not only associated with NGC 5252, but that its 
velocity field exhibits signs of rotation centered on the ULX.  If the 
gas velocities arise from virial motions, the ULX is plausibly associated with
a nuclear BH of an accreted dwarf galaxy.

Adopting the cosmological parameters $H_0 = 100 h = 67.8$ 
km~s$^{-1}$ Mpc$^{-1}$, $\Omega_m = 0.308$, and $\Omega_{\Lambda} = 0.692$ 
(\citealt{planck_2016}), the luminosity distance of NGC 5252 is 103 Mpc.  

\begin{figure*}[ht!]
\centering
\includegraphics[width=0.95\textwidth]{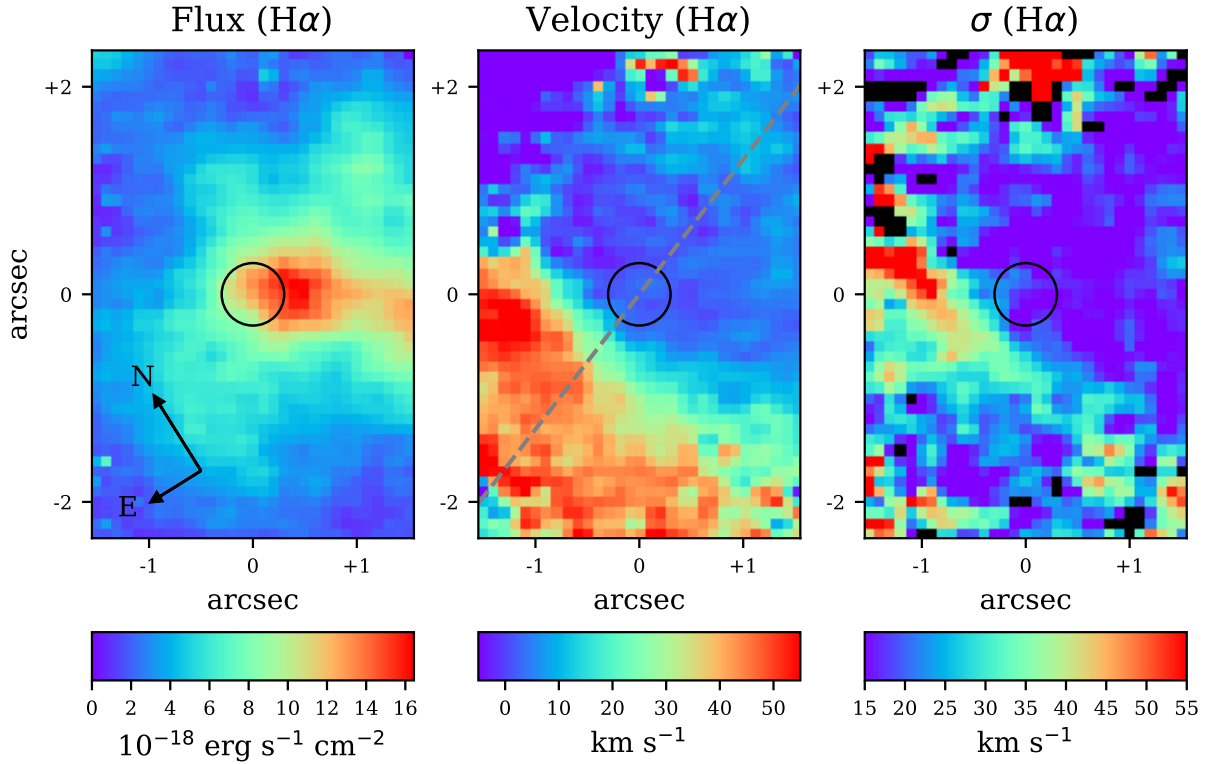}
\caption{
Distributions of \ha\ flux, velocity from the peak of \ha\ emission, and 
velocity dispersion of \ha\ emission. The velocity dispersion has been 
corrected for the instrumental resolution. The X-ray position of the ULX is 
denoted by the open circle, whose radius corresponds to the uncertainty 
of the position ($\sim$ 0\farcs3).  The dashed line in the middle panel 
marks the locus along which the position-velocity diagram in Figure 4 is 
extracted.
}
\end{figure*}
\vskip 0.1in

\begin{figure}[t!]
\centering
\includegraphics[width=0.47\textwidth]{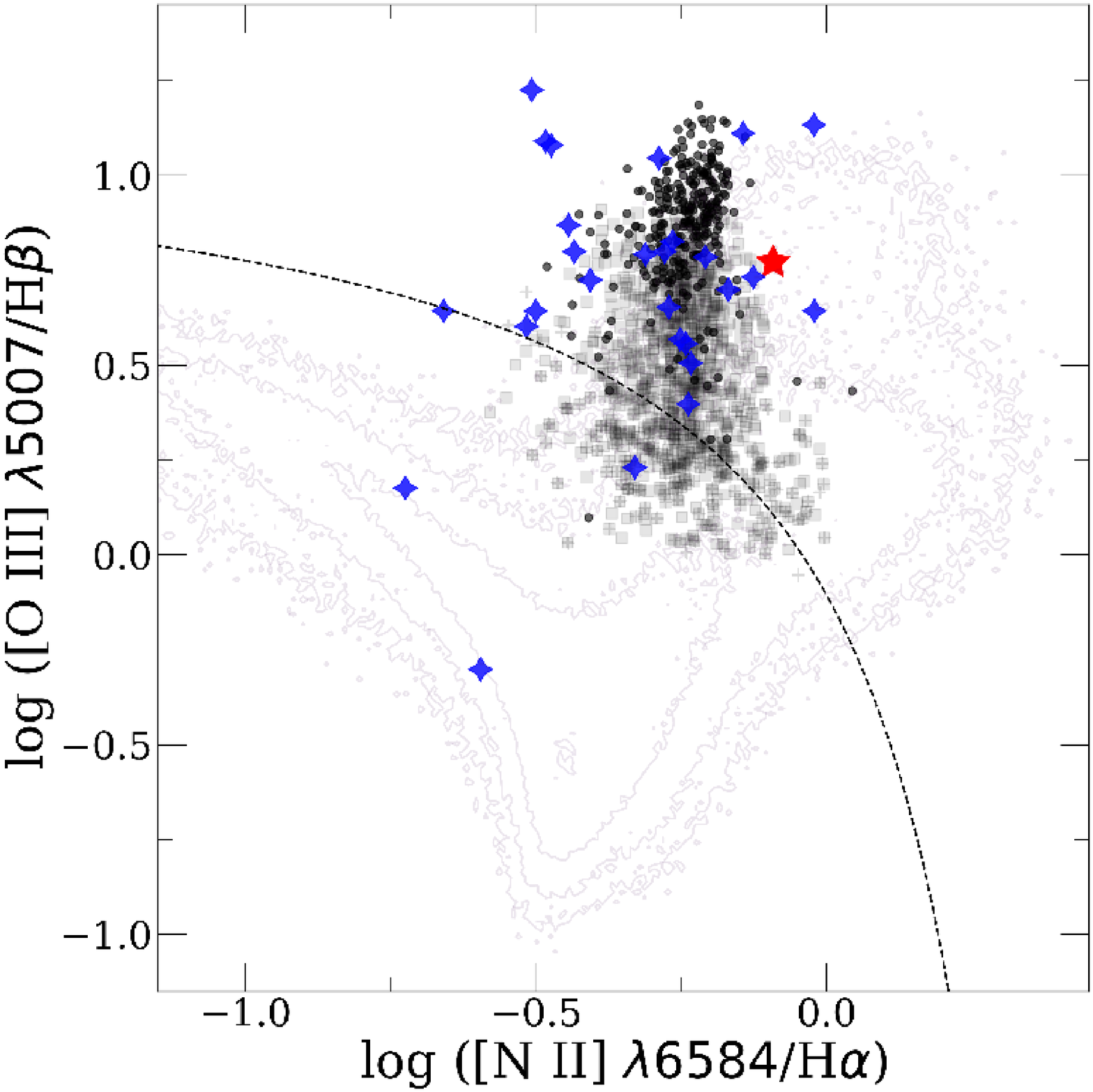}
\caption{
Optical line intensity ratio diagnostic diagram plotting 
\oiii\ $\lambda 5007$/H$\beta$ versus \nii\ $\lambda 6584$/H$\alpha$ 
(\citealt{baldwin_1981}).  Dark dots denote measurements of individual spaxels 
from our 
integral-field spectral observations; grey squares give upper limits on 
\nii/\ha, and grey crosses plot lower limits on \oiii/\hb.  Active galaxies 
with low-mass BHs ($\langle M_{\rm BH} \rangle = 3.5\times10^5$ \solmass; 
\citealt{ludwig_2012}) are shown as blue crosses.  The red star represents the 
integrated emission for the ULX host galaxy NGC 5252. The dashed line 
demarcates active galaxies from star-forming galaxies (\citealt{kewley_2001}). 
Background contour represents distribution for nearby galaxies from Sloan 
Digital Sky Survey (\citealt{kauffmann_2003}).}
\end{figure}
\vskip 0.1in

\section{Observations and Data Reduction}

We obtained integral-field unit spectra using the Gemini Multi-Object 
Spectrographs (GMOS) on the Gemini-north telescope on 2015 March 18. The data 
were taken with an R600 grating using a one-slit mode to cover \hb\ and \ha\ 
simultaneously, resulting in a field-of-view of 5\farcs0$\times$3\farcs5 
sampled by 500 elements.  The spectrum covers 4900--7700 \AA\ with a FWHM 
spectral resolution of $\sim$3700, as determined from the widths of the sky 
emission lines.  
The total on-source integration time was 3.75 hours.  The sky 
conditions were clear, and the seeing inferred from the guide camera was 
between 0\farcs1 and 0\farcs3. 

We reduced the data using the {\tt Gemini} package in IRAF. The flat-fielding 
was done using a twilight flat and flats taken with quartz halogen lamps, and
wavelength calibration was performed using CuAr arc images. We used the sky 
images taken from a blank region located 1$^\prime$ away from the science 
object for sky subtraction. We extracted the spectra in 0\farcs1 spatial 
elements and combined the spectra for each element. Finally, flux calibration 
was done using B9.6V star HR~5501 observed during the same night.

Although the signal-to-noise ratio ($S/N$) of the continuum in each spaxel
is $\leq 1$ on average, strong emission lines (e.g., \oiii$ \lambda$5007 and 
\ha) are detected with $S/N \geq 3$ in the majority of the spaxels (Fig. 1).
To investigate the properties of various emission lines (e.g., line width, 
flux, central wavelength), we fit the strong emission lines with a single 
Gaussian component. If the $S/N$ of the line is less than 3, we set upper 
limits based on 3 times the RMS noise in the associated continuum and using 
the line width measured from detected emission lines (\ha\ or \oiii).        

\begin{figure*}[t]
\includegraphics[width=0.95\textwidth]{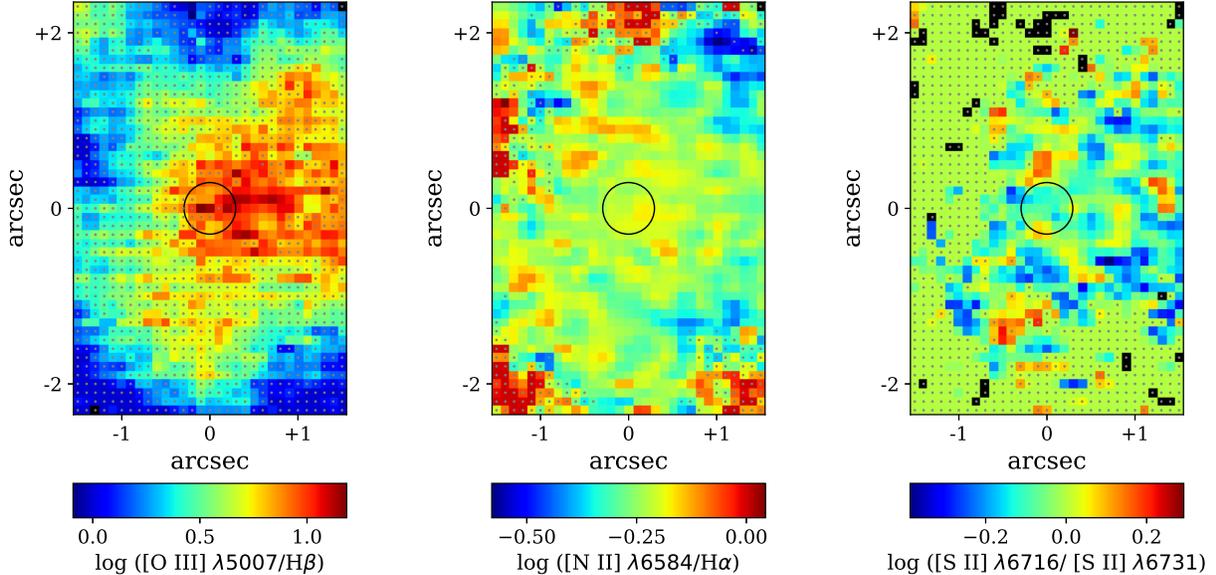}
\caption{
Distribution of flux intensity ratios for \oiii\ $\lambda$5007/\hb,
\nii\ $\lambda$6584/\ha, and \sii\ \lamb6716/\sii\ \lamb6731). Small dots 
denote lower limits (left), upper limits (middle), and unknown values (right),
respectively.  The position of the ULX is denoted by the open circle, whose 
radius corresponds to the uncertainty of the position ($\sim$ 0\farcs3). 
}
\end{figure*}

\section{Spectral Properties}
\subsection{Rotational Feature}
Figure 1 presents distributions of line flux, velocity measured from the peak 
of the \ha\ emission line, and its associated velocity dispersion ($\sigma$),
corrected for instrumental resolution.  The distribution of \ha\ flux clearly 
shows that the ULX is coincident with the ionized gas (\citealt{kim_2015}). 
While the structure of the ionized gas appears to be somewhat asymmetric 
relative to the ULX, the X-ray position{\footnote{The uncertainty of the 
position is $\sim0\farcs3$.}} of the ULX agrees well with one of the 
peak positions of the \ha\ flux (see also Fig. 4 in \citealt{kim_2015}). 
Note that the kinematical and morphological distributions of \oiii\ emission 
are very similar to those of \ha. 

More interstingly, the spatial distribution of the peak velocity of \ha\ 
exhibits rotational motion. This is more clearly seen in the position-velocity 
diagram along the axis denoted by the dashed line in Figure 1 (\S{4}).  One 
can argue that shocks may be responsible for the velocity gradient. If this 
were the case, the velocity dispersion should increase substantially at the 
shock front. Although, the distribution of $\sigma$(\ha) shows a very mild
increase near the position of the velocity gradient, such a pattern is not 
present in \oiii, which should be even more sensitive to shocks than \ha\ 
(Fig. 2). 
In addition, if shocks contribute to the ionization, we expect relatively low 
values of \oiii/\hb\ ($\lax\ 3$) and \nii/\ha\ ($\lax\ 3$) compared to AGNs, 
and relatively high velocity dispersions (often larger than 100 \kms; Ho et 
al. 2014). In our case, however, \oiii/\hb\ tends to be greater than 3 
and the velocity dispersions of \ha\ are less than 55 \kms\ (Figs. 1 and 2).
Thus, shocks seem 
unlikely to be important.  
And while \oiii\ can effectively trace outflows, the lakc of  excess velocity 
dispersion in the vicinity of the ULX suggests that 
outflows are not significant.

It is worth noting that the center of rotation is offset by $\sim$0\farcs5 
($\sim230$ pc) with respect to position of the ULX.  This indicates that either
the surrounding gas might be gravitationally unbound to the ULX or the gas was 
decoupled from the nucleus, possibly during the merger, if the ULX was 
associated with a merging galaxy (see \S{4}). The spatial distribution of \ha\ 
shows that the ionized gas extends up to 3\asec\ in diameter ($\sim1.4$ kpc) 
and is highly disturbed.

\subsection{Source of the Ionization}
The \oiii\ $\lambda 5007$/H$\beta$ versus \nii\ $\lambda 6584$/H$\alpha$ 
diagnostic diagram (\citealt{baldwin_1981}) clearly shows that the gas 
ionization 
is mostly dominated by an AGN rather than by star formation (Fig. 2).  
\oiii/\hb\ 
reaches a maximum at the position of the ULX (Fig. 3).   This flux ratio 
increases mainly with increasing ionization parameter, metallicity, or electron
density (e.g., \citealt{ferland_1983}).  
However, \sii\ \lamb6716/\sii\ \lamb6731 shows no evidence of 
a rise in electron density in the vicinity of the ULX, nor is there any 
sign for a higher metallicity 
from \nii\ $\lambda 6584$/\ha\ (e.g., \citealt{storchi_1989}).  Altogether,
it appears that the elevated \oiii/\hb\ associated with the ULX arises from 
an increase in ionization parameter, and hence that the ULX itself is the main 
source of ionization. 

\section{The Physical Association of the ULX with NGC 5252}

In view of the complexity of the gas distribution in the host galaxy NGC 5252
(\citealt{prieto_1996}), we cannot completely rule out the possibility that the
ULX is a background AGN interloper.  \citet{kim_2015} argued against this 
interpretation due to the lack of emission lines that should be associated 
with such a background source, a result reaffirmed in the current set of GMOS 
observations.  They also rejected the possibility that the source is a 
background BL Lac object (a class of AGNs with very weak or no emission lines) 
based on the broad-band spectral energy distribution of the source and the 
absence of X-ray and radio variability.

The observations presented here further strengthen the proposal that the ULX 
resides within NGC 5252.  First, the position-velocity diagram indicates that 
the ionized gas projected toward the immediate vicinity of the ULX shows 
{\it systematic}\ motions plausibly indicative of rotation.  The rotational 
velocity reaches a maximum of $\sim 20$ \kms, but it is offset by 
$r\approx0\farcs5$ ($\sim$230 pc; Fig. 4) from the position of the ULX.  With 
an assumed inclination angle of 45\deg, the dynamical mass is $M_{\rm dyn}
\approx 4\times10^7$ \solmass.  Although this mass resembles that of the most 
massive giant molecular clouds, its relatively large size and 
rotation-dominated kinematics place it apart from typical giant molecular 
clouds (e.g., \citealt{hughes_2010}). Instead, the velocity gradient more 
closely resembles 
that of the circumnuclear region of a dwarf galaxy (e.g., 
\citealt{oh_2015}). 

Furthermore, as discussed in \S{3}, the excitation source for the line emission
seems to coincide closely with the position of the ULX, strongly suggesting 
that the source of ionization for the gas is the ULX itself.  This causally 
ties the ULX to the gas physically residing within NGC 5252.   Moreover, 
$M_{\rm dyn}$ can be regarded as a strict upper limit on the mass of the ULX.  
This value of $M_{\rm BH}$ falls within the (admittedly poorly constrained) 
range of $10^{3.2} - 10^9$ \solmass\ previously estimated by \citet{kim_2015} 
based on the \oiii\ luminosity and the so-called BH fundamental plane (e.g., 
\citealt{merloni_2003}).  Taken at face value, the upper limit on $M_{\rm BH}$ 
indicates that this ULX can still be regarded as a possible candidate for an 
IMBH, although it depends on the (unknown) inclination angle.  

Nevertheless, we must exercise caution in interpreting the gas kinematics, 
since the velocity structure surrounding the ULX might be associated with the 
circumnuclear region of a merging dwarf galaxy.  The positional offset of the 
ULX with respect to the center of rotation may be a manifestation of tidal 
distortion during the infall of the progenitor galaxy.  If so, the true BH 
mass associated with the ULX may be much smaller than $M_{\rm dyn}$.  An even 
more extreme possibility is that, even if the ULX does reside within NGC 5252, 
it may just be passing by coincidentally and ionizing the gas in the cloud
(\citealt{krolik_2004}).  
In this case, the gas kinematics give no constraints on the mass of the ULX. 
However, we note that the positional uncertainty of the Chandra data can 
be as large as 0\farcs7 (90 percent confidence) calculated from the direct 
comparison between Chandra X-ray positions and those of optical 
counterparts{\footnote{\url{http://cxc.cfa.harvard.edu/cal/ASPECT/celmon/}}}. 
Thus, 
it is possible that the apparent positional offset is not significant.

If the ULX and its surrounding nebula indeed originate from the accretion of 
a dwarf galaxy by NGC 5252, then we expect the metallicity of the gas to be 
rather low, reflecting its origin in the low-mass progenitor 
(\citealt{tremonti_2004}).  The gas-phase metallicity can be estimated from 
the line-intensity ratio \nii/\ha\ (e.g., \citealt{pettini_2004}; 
\citealt{groves_2006}). Consistent 
with this expectation, Figure 2 illustrates that the ionized gas around the 
ULX has significantly lower \nii/\ha\ than the main body of the more massive 
host galaxy NGC 5252.  The spatially resolved emission surrounding the ULX 
overlaps with the locus of low-mass active galaxies having $\langle M_{\rm BH} 
\rangle = 3.5\times10^5$ \solmass\ (\citealt{greene_2007}; 
\citealt{ludwig_2012}). This result
strongly supports our hypothesis that a merging dwarf galaxy 
delivered the ULX to NGC 5252, and, indirectly, that the BH associated with 
the ULX has a low BH mass.

\acknowledgements
We are grateful to an anonymous referee for fruitful comments. MK thanks
Jae-Joon Lee for useful discussions.
This research was supported by the Basic Science Research Program through the 
National Research Foundation of Korea (NRF) funded by the Ministry of 
Science, ICT \& Future Planning (No. NRF-2017R1C1B2002879) and by K-GMT 
Science Program (PID: GN-2015A-Q-201) of Korea Astronomy 
\begin{figure}[th]
\centering
\includegraphics[width=0.47\textwidth, bb= 90 180 492 542]{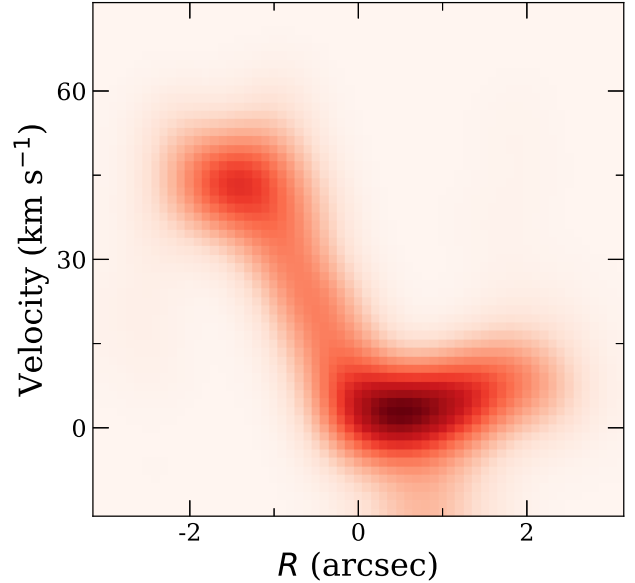}
\caption{
Position-velocity diagram of \ha\ emission. The position is determined along 
the dashed line in Figure 1. The velocity is defined relative to the 
systematic velocity of the ULX (\citealt{kim_2015}).
}
\end{figure}
\vskip 1.0in
\noindent
and Space Science 
Institute (KASI).  LCH was supported by 
the National Key Program for Science 
and Technology Research and Development (2016YFA0400702) and the National 
Science Foundation of China (11473002).  MI was supported by NRF grant
No. 2017R1A3A3001362, funded by the Korea government (MSIP).


\end{document}